\documentclass[]{aa}  

\usepackage{graphicx}
\usepackage{txfonts}
\begin{document}

   \title{APEX-SEPIA660 Early Science:\\ Gas at densities above $10^7$~cm$^{-3}$ towards OMC-1
   	\thanks{This publication is based on data acquired with the Atacama Pathfinder Experiment (APEX). APEX is a collaboration between the Max-Planck-Institut fur Radioastronomie, the European Southern Observatory, and the Onsala Space Observatory}
   }
   

\author{A. Hacar
          \inst{1,2}
          \and
          M. R. Hogerheijde \inst{1,3}
          \and
          D. Harsono \inst{4}
          \and
          S. Portegies Zwart \inst{1}
          \and
          C. De Breuck \inst{5}
          \and 
          K. Torstensson  \inst{5}
          \and
          W. Boland \inst{6}
          \and
          A. M. Baryshev \inst{6}
          \and
          R. Hesper \inst{6}
          \and
          J. Barkhof \inst{6}
          \and
           J. Adema \inst{6}
           \and
           M. E. Bekema \inst{6}
           \and
           A. Koops \inst{6}
           \and
           A. Khudchenko \inst{6}
           \and
           R. Stark \inst{7}
          }

\institute{
         	   Leiden Observatory, Leiden University, P.O. Box 9513, 2300-RA Leiden, The Netherlands\\
         	   \email{hacar@strw.leidenuniv.nl} 
         \and
         University of Vienna, Department of Astrophysics, T\"urkenschanzstrasse 17, 1180 Vienna, Austria
         \and
         Anton Pannekoek Institute for Astonomy, University of Amsterdam, the Netherlands
	   	  \and
		  Institute of Astronomy and Astrophysics, Academia Sinica, No.1, Sec. 4, Roosevelt Rd, Taipei 10617, Taiwan, R.O.C.
		  \and
                  European Southern Observatory, Karl Schwarzschild Strasse 2, 85748 Garching, Germany
                  \and
                  Netherlands Research School for Astronomy (NOVA), Kapteyn Astronomical Institute, Landleven 12, 9747 AD Groningen,
The Netherlands
				\and
				Netherlands Research School for Astronomy (NOVA), Leiden Observatory, Leiden University, P.O. Box 9513, 2300-RA Leiden, The Netherlands
             }

   \date{---}

 
  \abstract
{The star formation rates and stellar densities found in young massive clusters suggest that these stellar systems originate from gas at densities n(H$_2$)~$>10^6$~cm$^{-3}$. Until today, however, the physical characterization of this ultra high density material remains largely unconstrained in observations. 
}
{We investigated the density properties of the star-forming gas in the OMC-1 region located in the vicinity of the Orion Nebula Cluster (ONC).}
{We mapped the molecular emission at 652~GHz in OMC-1 as part of the APEX-SEPIA660 Early Science.}
{We detect bright and extended N$_2$H$^+$~(J=7--6) line emission along the entire OMC-1 region.
Comparisons with previous ALMA data of the (J=1--0) transition and radiative transfer models indicate that the line intensities observed in this N$_2$H$^+$~(7--6) line are produced by large mass reservoirs of gas at densities n(H$_2$)~$>10^7$~cm$^{-3}$.}
{The first detection of this N$_2$H$^+$~(7--6) line at parsec-scales demonstrates the extreme density conditions of the star-forming gas in young massive clusters such as the ONC.
Our results highlight the unique combination of sensitivity and mapping capabilities of the new SEPIA660 receiver for the study of the ISM properties at high frequencies.}



   \maketitle
%

\section{Initial gas conditions in massive clusters}

Investigating the origin of massive stellar clusters \citep[$>10^4$~M$_\sun$,][]{SPZ10} in the Milky Way is of paramount importance to understand the star-formation process across Cosmic Times \citep{KRU19}.
Massive clusters represent the most extreme examples of star-formation and the local analogues for the gas conditions at high redshifts \citep[see][]{LON14} .
Usually located at kpc distances (e.g. towards the Galactic Centre), the characterization of these massive clusters is limited by sensitivity and resolution effects. Massive clusters form inside highly extincted gas clumps that can only be scrutinized using radio interferometric observations \citep[e.g.][]{GIN18}.
The molecular material in these massive clusters is also quickly disrupted by the strong stellar feedback generated by their active stellar populations. 
As result, the initial gas conditions inside massive clusters remain largely unconstrained in observations.

Stellar densities of n(H$_2$)~$> 10^6$~cm$^{-3}$ (or  $>10^7$~M$_\sun$/pc$^3$)
have been identified in massive clusters \citep{POR04}, which suggests that these clusters originated from gas cloud at even higher densities.
Different studies indicate that  the molecular precursors of these massive clusters in the Milky Way may present average gas densities of n(H$_2$)~$\sim10^4$~cm$^{-3}$ \citep[e.g.][]{KAU17}.
Super star clusters, such as Arches, are expected to be originated from gas clumps at densities above n(H$_2$)~$>10^6$~cm$^{-3}$ \citep{WAL15}.
These gas densities largely exceed the mean values observed in nearby clouds predicting dramatically different free-fall times ($\tau_{ff}\propto \mathrm{n(H}_2\mathrm{)}^{-1/2}$) and star-formation rates (SFR~$\propto \mathrm{n(H}_2\mathrm{)}^{1/2}$) between these regions \citep[see][]{KRU12}.
However, and while extreme density values are reported for compact hot cores ($<$~0.1~pc) in massive environments \citep[e.g.][]{GEN82},
the detection of gas at densities above n(H$_2$)~$\gtrsim 10^7$~cm$^{-3}$ at parsec-scales remains elusive.

\begin{figure*}[!h]
	\centering
	\includegraphics[width=\textwidth]{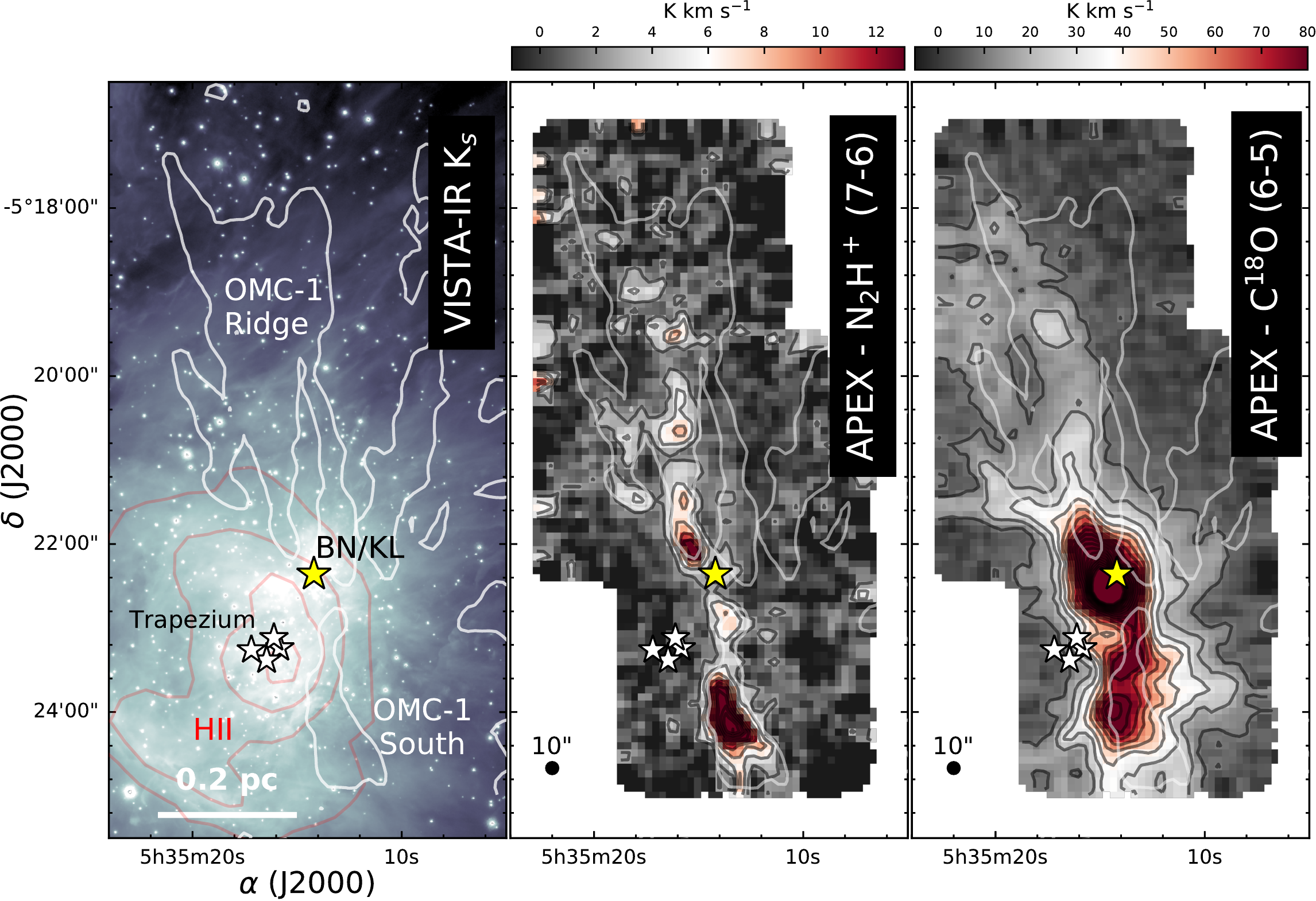}
	\caption{
		New SEPIA660 observations along the OMC-1 region. From left to right:
		{\bf (a)} VISTA-IR K$_s$ band \citep{MEI16},
		{\bf (b)} APEX-SEPIA660 N$_2$H$^+$~(7--6),
		and {\bf (c)} APEX-SEPIA660 C$^{18}$O (6--5) maps (this work).
		All molecular maps are convolved into a common Nyquist grid with a final resolution of 10".
		The intensity the H41$\alpha$ emission tracing the extension of the ONC HII nebula \citep[red contours,][]{HAC19} is indicated in the VISTA image.
		Panels b and c display equally spaced contours (black) every $W(\mathrm{N}_2\mathrm{H}^+ (7-6))=$~2.5~K~km~s$^{-1}$ and $W(\mathrm{C}^{18}\mathrm{O}\:  (6-5))=$~10~K~km~s$^{-1}$,  respectively.
		For reference, the position of the Trapezium (white stars) and the Orion BN source (yellow star), as well as the first contour of the N$_2$H$^+$~(1--0) emission
		($W(\mathrm{N}_2\mathrm{H}^+ (1-0))=$~1.0~K~km~s$^{-1}$,white contour) \citep[][see also Fig.~\ref{fig:ratios}a]{HAC18}, are displayed in all panels.
	}
	\label{fig:observations}
\end{figure*}
\begin{figure*}[!ht]
	\centering
	\includegraphics[width=\textwidth]{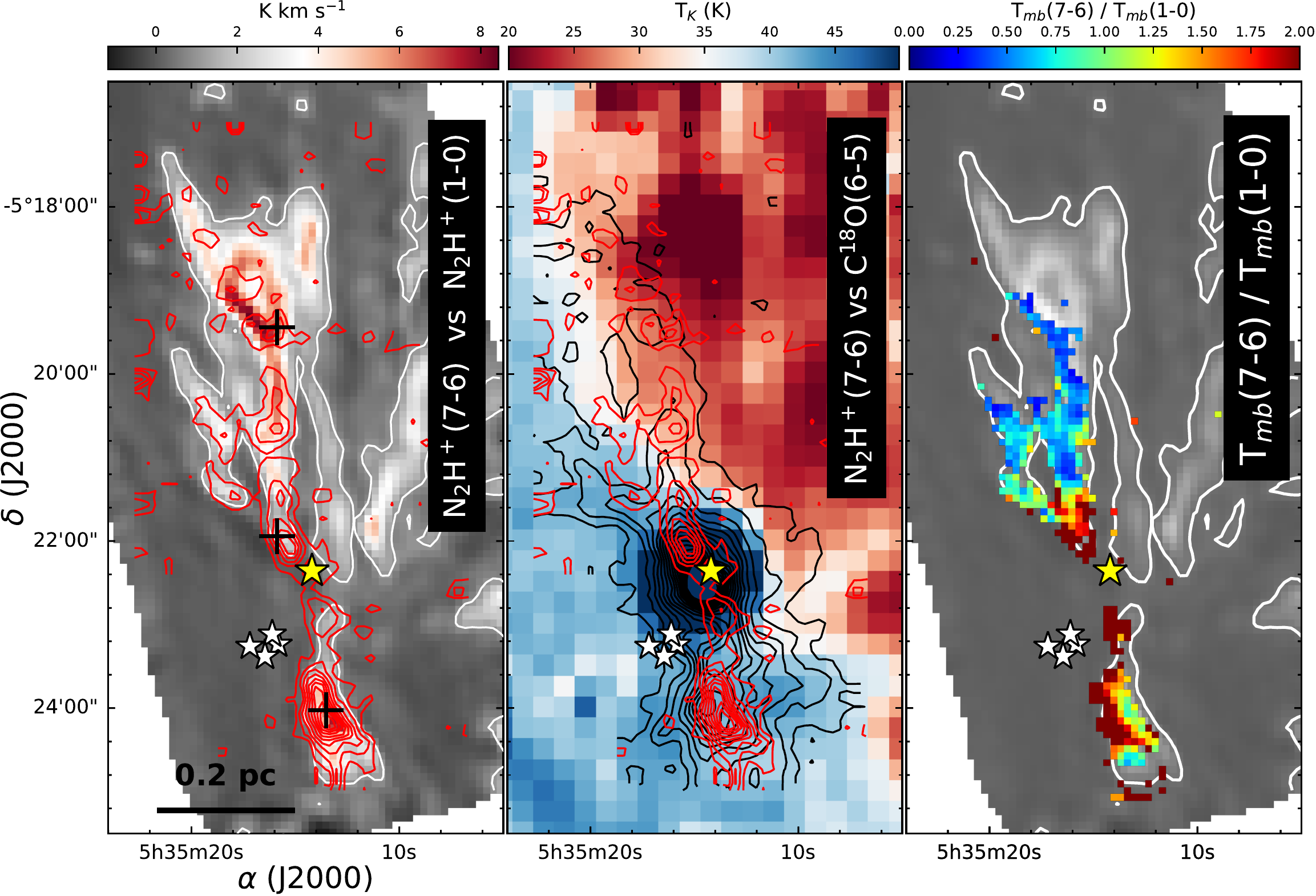}
	\caption{
		From left to right: 
		{\bf (a)} Distribution of the total integrated N$_2$H$^+$ (7--6) emission detected by APEX-SEPIA660 (red contours; this work) superposed to the N$_2$H$^+$ (1--0) emission observed by ALMA \citep[colour scale;][]{HAC18} both convolved into a resolution of 10".
		{\bf (b)} Comparison between N$_2$H$^+$ (7--6) (red contours) and 
		C$^{18}$O (6--5) (black contours) emission as function of the gas kinetic temperature \citep[colour scale,][]{HAC19}.
		Both N$_2$H$^+$ and C$^{18}$O contour levels are similar to those displayed in Fig.~\ref{fig:observations}.
		{\bf (c)} $\frac{T_\mathrm{mb}(\mathrm{N}_2\mathrm{H}^+\mathrm{(7-6)})}{T_\mathrm{mb}(\mathrm{N}_2\mathrm{H}^+\mathrm{(1-0)})}$ line peak ratio (colour scale) plotted over the total integrated intensity N$_2$H$^+$ (1--0) (grey scale).
		For reference, the position of the Trapezium (white stars) and the Orion BN source (yellow star) are displayed in all maps.  The first contour of the N$_2$H$^+$~(1--0) emission ($W(\mathrm{N}_2\mathrm{H}^+ (1-0))=$1.0~K~km~s$^{-1}$,white contour) is also indicated in panels a \& c.
		The black crosses in panel a indicate the position of the representative spectra shown in Figure~\ref{fig:spectra}.
	}
	\label{fig:ratios}
\end{figure*}

The increasing sensitivity of both single-dish and interferometers
has popularized the use of N$_2$H$^+$ as density selective tracer in star formation studies  \citep[e.g.][]{CAS02}.
A combination of excitation and chemical effects (critical density, abundance, and depletion) enhances the emission of this N-bearing molecule in dense environments ($> 10^4$~cm$^{-3}$) with respect to other standard cloud tracers (e.g. CO, HCN, or HCO$^+$) commonly biased towards lower density ($\sim 10^3$~cm$^{-3}$) and/or warm material \citep[$T_\mathrm{K}>20$~K, e.g.][]{PET17}. 
The relatively higher abundances of N$_2$H$^+$ also makes this molecule a favourable target for observations  in comparison with other deuterated isotopologues (e.g. N$_2$D$^+$) and species (e.g. DCO$^+$).
Most studies typically investigate low frequency transitions of this molecule, such as 
N$_2$H$^+$ (1--0) \citep[93~GHz; e.g.][]{HAC18}
or N$_2$H$^+$ (3--2) \citep[279~GHz; e.g.][]{TEN20}, limiting the dynamic range of these observations to densities of n(H$_2$)~$\lesssim 10^6$~cm$^{-3}$. 
However, the observation of high-J transitions (J$>$4--3) necessary to confirm the existence of gas at higher densities has been largely hampered by the more challenging access to frequencies above $>$300~GHz.

In this paper we report the first detection of extended N$_2$H$^+$~(J=7--6) emission at $\sim$652~GHz in the vicinity of the Orion Nebula Cluster (ONC) mapped with the new SEPIA660 heterodyne receiver (i.e. ALMA Band 9) recently installed at the APEX-12m radiotelescope (Sect.~\ref{sec:observations}).
The widespread detection of bright  N$_2$H$^+$ (7--6) emission at scales of $>$~0.5~pc demonstrates the presence of gas at densities  n(H$_2$)~$>10^7$~cm$^{-3}$ (Sect.~\ref{sec:nebula} \& \ref{sec:highdensity}).
Our new SEPIA660 observations provide us with the first direct evidence of the presence of large gas reservoirs at extremely high densities during the early evolution of young massive clusters  (Sect.~\ref{sec:final}).

\section{New SEPIA660 observations}\label{sec:observations}

The ONC is the nearest high-mass star-forming region \citep[D=414pc,][]{MEN07} regularly used as local template for cluster studies. 
The ONC is partially embedded in the OMC-1 region, an active star-forming  cloud including the Orion BN/KL region, widely investigated in the past at large-scales using both millimeter single-dish {\citep[e.g.][]{ UNG97} } and interferometric observations \citep[e.g.][]{WIS98}.
Highlighting the filamentary nature of this OMC-1 cloud \citep{MPI90}, recent ALMA (Band 3) observations of N$_2$H$^+$~(J=1--0) revealed the intrinsic structure of the star-forming gas in this region forming a complex networks of narrow fibers \citep{HAC18}.
Continuum \citep[e.g.][]{TEI16} and line measurements \citep{HAC18} in this massive cloud suggest that these fiber structures may present densities significantly larger than those found in low-mass environments \citep[see][for a discussion]{HAC18}.
A detailed analysis of the N$_2$H$^+$~(1--0) hyperfine line opacities and excitation suggest that the density of these fibers reaches values of n(H$_2$)~$> 10^7$~cm$^{-3}$ \citep[see][for a discussion]{HAC18}. This hypothesis is supported by the recent detection of extended  N$_2$H$^+$~(3--2) emission towards the entire OMC-1 region \citep{TEN20}.

We have mapped the central part of the OMC-1 region with the new Swedish-ESO PI instrument for APEX (SEPIA) \citep{BEL18}  installed at the APEX-12m telescope in Chajnantor (Chile). 
Our observations used the new SEPIA660 detector, a dual polarization 2SB receiver operating between 578 and 738~GHz (similar to an ALMA Band 9 receiver) developed by the Netherlands Research School for Astronomy (NOVA) instrumentation group at the Kapteyn Astronomical Institute in Groningen (The Netherlands) \citep{SEPIApaper}.
Our maps cover the entire OMC-1 region (see Fig.~\ref{fig:observations}), from its northern OMC-1 Ridge to the Orion South proto-clusters,  including the Orion BN/KL region (see labels in Fig.~\ref{fig:observations}a). 
We combine four on-the-fly (OTF) maps covering a total area of $\sim 450 \times 200$ arcsec$^2$, or approximately $\sim 0.9 \times 0.4$ pc$^2$ in size at the distance of Orion. 
Each OTF submap, with a typical area of 150$\times$150 arcsec$^2$, was obtained in Position-Switching mode and was executed multiple times combining orthogonal coverages. Our observations were carried out in August 2019 under excellent weather conditions with a Precipitable Water Vapor (PWV) of PWV$\le$~0.5~mm as part of the NOVA Guaranteed Time (ESO Proj. ID: E-0104.C-0578A-2019)\footnote{
	This work is a continuation of the ORION-4D project (PI: A. Hacar). See more information in \url{https://sites.google.com/site/orion4dproject} .}.

Our study targets two specific lines, namely, N$_2$H$^+$~(J=7--6) (652095.865~MHz) and C$^{18}$O~(J=6--5) (658553.278~MHz) \citep[CDMS and VAMDC databases,][]{CDMS2,CDMS}, observed simultaneously with a native spectral resolution of 240 kHz (or 0.11 km~s$^{-1}$) thanks to the large instantaneous bandwidth (8~GHz) of the new SEPIA660 receiver connected to an XFFTS backend\footnote{For each 4~GHz subband, the XFFTS backend installed at APEX has a maximum resolution of 65536 channels (or 61~kHz per channel). In order to reduce the data rate in our high cadence OTF-maps, our observations reduced the effective spectral resolution down of this XFFTS backend to 16384 channels (or  240~kHz) per sideband and polarisation.}. Each molecular species was extracted, reduced, and combined independently using the software GILDAS/CLASS. We convolved our N$_2$H$^+$~(7--6) and C$^{18}$O (6--5) datasets into a uniform Nyquist sampled grid with a final resolution of 10~arcsec. After that, each individual spectrum {\bf was} baseline subtracted and calibrated into main beam temperature units (T$_{mb}$) assuming a typical main-beam efficiency of $\eta_{mb}=0.4$ measured in Uranus\footnote{http://www.apex-telescope.org/telescope/efficiency}. 

We display the total integrated intensity of our new SEPIA660 observations along the OMC-1 region in Figure~\ref{fig:observations}. Remarkably, we clearly detect extended N$_2$H$^+$~(7--6) emission  with a total integrated intensity $W(\mathrm{N}_2\mathrm{H}^+)>$~2.5~K~km~s$^{-1}$ showing different cores and fibers along the whole extension of our maps (Fig.~\ref{fig:observations}b). Extremely bright emission peaks, exceeding values of $W(\mathrm{N}_2\mathrm{H}^+)>$~8~K~km~s$^{-1}$, are identified  at the north of the Orion BN/KL region as well as the Orion South proto-cluster.
Additional emission peaks in this N$_2$H$^+$~(7--6) transition, typically with $W(\mathrm{N}_2\mathrm{H}^+)\sim$5~K~km~s$^{-1}$,  are found coincident with similar concentrations detected in N$_2$H$^+$~(1-0) by ALMA (white contours in Fig.~\ref{fig:observations}; see also Fig.~\ref{fig:ratios}a). Several of these peaks appear to be connected by a more diffuse N$_2$H$^+$~(7--6) emission extending North-South along the main spine of the OMC-1 region. Overall, the N$_2$H$^+$~(7--6) emission is restricted to regions containing high column density material traced in the continuum above N(H$_2$)~$>10^{22}$~cm$^{-2}$ \citep[e.g.][]{LOM14}.

In contrast, our C$^{18}$O (6--5) map shows a clear radial distribution in emission centred at the position of the Orion BN/KL region (Fig.~\ref{fig:observations}c). At its central peak, the C$^{18}$O (6--5) emission reaches values of $W$(C$^{18}$O )$>$100~K~km~s$^{-1}$, including prominent line wings associated to the Orion BN/KL outflow \citep{BAL11}.  Bright C$^{18}$O (6--5) emission, with W(C$^{18}$O )$>$50~K~km~s$^{-1}$, is also observed towards the southern end of the OMC-1 Ridge as well as the northern half of the OMC-1 South proto-cluster. More diffuse but still clearly detected, this emission continues at large scales below N(H$_2$)~$\sim10^{21}$~cm$^{-2}$, tracing the warm molecular material around the Orion Nebula in agreement with similar $^{12}$CO (J=4--3) \citep{ISH16} and [CII] \citep{PAB19} observations in this region. Moreover, the C$^{18}$O (6--5) emission shows an arc-like structure tracing the edge of the HII region previously detected in H recombination lines \citep[e.g.][]{HAC19}. These properties denote
the preference of this C$^{18}$O (6--5) transition to trace low column density material directly illuminated by the Orion cluster.

\section{Dense gas exposed to the ONC}\label{sec:nebula}

The observed anticorrelation between N$_2$H$^+$ and C$^{18}$O is generated as part of the chemical evolution of the gas during the collapse and formation of stars in molecular clouds \citep[see][and reference therein]{BER07}. 
	As part of Nitrogen chemistry, N$_2$H$^+$ is directly formed from N$_2$ via:
	\begin{equation}\label{f1}
		\mathrm{H}_3^+ + 	\mathrm{N}_2 \rightarrow \mathrm{N}_2\mathrm{H}^+ + 	\mathrm{H}_2
	\end{equation}
	in direct competition with CO:
		\begin{equation}\label{f2}
	\mathrm{H}_3^+ + 	\mathrm{CO} \rightarrow \mathrm{HCO}^+ + 	\mathrm{H}_2 .
	\end{equation}
On the other hand, N$_2$H$^+$ is destroyed via proton transfer with CO:
	\begin{equation}\label{d1}
\mathrm{N}_2\mathrm{H}^+ + 	\mathrm{CO} \rightarrow \mathrm{HCO}^+ + \mathrm{N}_2
\end{equation}
as well as via dissociative recombination:
\begin{equation}\label{d2}
\mathrm{N}_2\mathrm{H}^+ + 	\mathrm{e} \rightarrow \mathrm{NH} + \mathrm{H}\; \mathrm{or}\; \mathrm{N}_2 + \mathrm{H}.
\end{equation}
 \citep[see][for a full discussion]{AIK15,vHOF17}.
At the standard low-densities (n(H$_2$)~$\gtrsim 10^2$~cm$^{-3}$) and cold temperatures ($T_\mathrm{K}\sim$~10~K) of the ISM the  efficient formation of N$_2$H$^+$ (reaction \ref{f1}) is inhibited by the presence of large amounts of CO in the gas phase (reactions \ref{f2} and \ref{d1}). 
The abundance of N$_2$H$^+$ is rapidly enhanced during the gas collapse after the depletion of CO once this latter species is frozen onto the dust grains at densities above n(H$_2$)~$\gtrsim 10^4$~cm$^{-3}$. N$_2$H$^+$ is later destroyed once the CO is evaporated from the dust heated at temperatures above $T_\mathrm{K}=$~20~K (reaction \ref{d1}) or under the presence of free electrons (reaction \ref{d2}) produced  in HII regions such as the Orion Nebula. These selective properties make N$_2$H$^+$ an ideal tracer of the dense and cold material in molecular clouds (n(H$_2$)~$\gtrsim 10^5$~cm$^{-3}$, $T_\mathrm{K}\lesssim$~30~K). 
 Interestingly, while the N$_2$H$^+$ formation rate is almost independent of temperature  (reaction \ref{f1}), the rate of dissociative recombinations (reaction \ref{d2}) decreases at increasing temperatures \citep[e.g.][]{VIG12} reducing the destruction rate of this molecule at high temperatures.

The excitation conditions of the high-J N$_2$H$^+$ and C$^{18}$O transitions obtained in our SEPIA660 observations enhance the chemical differences of these tracers\footnote{Critical densities calculated using the latest LAMDA collision coefficients for both para- and orto-H$_2$ for C$^{18}$O and total H$_2$ in the case of N$_2$H$^+$.}. Due to its high dipole moment ($\mu=3.4$~D), the N$_2$H$^+$  (J=7--6) (E$_{u}=125$~K) line can only be collisionally excited at densities comparable to its critical density
 n$_{crit}\left(\mathrm{N}_2\mathrm{H}^+\:(7-6)\right)\sim 5\times 10^7$~cm$^{-3}$. In contrast, the lower dipole moment of C$^{18}$O ($\mu=0.11$~D) reduces the critical density of similarly high-J transitions, such as the C$^{18}$O (J=6--5) line, down to n$_{crit}(\mathrm{C}^{18}\mathrm{O} \: (6-5))\sim 5\times 10^5$~cm$^{-3}$. 
 The reduction of the effective critical density at high column densities and temperatures \citep[see ][for a full discussion]{SHI15}, similar to those found in the OMC-1 region, make the  C$^{18}$O
(J=6--5) transition (E$_{u}=110$~K) 
sensitive to more diffuse ($\lesssim 10^4$~cm$^{-3}$) and typically hotter material ( $T_\mathrm{K}>$~60~K).
The different gas regimes traced by these two molecules can be seen in previous observations of lower-J transitions in Orion such as N$_2$H$^+$  (J=1--0) \citep[e.g.][]{HAC18} and C$^{18}$O  (J=1--0) \citep[e.g.][]{KON18}.

The different gas regimes traced by the N$_2$H$^+$~(7--6) and C$^{18}$O (6--5) lines become apparent by the complementary distribution of their emission maps shown in Figure~\ref{fig:ratios}. Similar to the (1--0) transition (Fig.~\ref{fig:ratios}a), we find no significant N$_2$H$^+$~(7--6) emission at the position of the hot Orion BN/KL region \citep[see also][]{SCH01}, coincident with emission peak of the C$^{18}$O (6--5) line (Fig.~\ref{fig:ratios}b). We observe this same behaviour at the bright edge of the HII nebula devoid of N$_2$H$^+$~(7--6) {emission but well delineated by } brighter C$^{18}$O (6--5) emission. 
On the other hand, most of the N$_2$H$^+$~(7--6) peaks, as well as of the (1--0) emission, are located in well shielded and typically colder regions with low or no significant C$^{18}$O (6--5) detections (see OMC-1 Ridge region in Fig.~\ref{fig:ratios}b). The OMC-1 South protocluster appears as the only exception to this general behaviour. The extraordinary conditions of this cluster, engulfed by the Orion Nebula and seen face-on along the line-of-sight \citep{ODE01}, can explain the bright emission of both molecules in this region.

Although not coincident, many of the N$_2$H$^+$~(7--6) emission peaks are indeed located next to regions with enhanced C$^{18}$O (6--5) emission (see different contours in Fig.~\ref{fig:ratios}b). This systematic shift is particularly visible along the OMC-1 Ridge, where multiple clumps detected in N$_2$H$^+$ seem to be illuminated in C$^{18}$O in the direction of the ONC. 
These results suggest that the detected N$_2$H$^+$~(7--6) emission may be showing the densest molecular gas in the OMC-1 region in the close proximity to the edge of the Orion Nebula.

\begin{figure*}[!ht]
	\centering
	\includegraphics[width=1.0\textwidth]{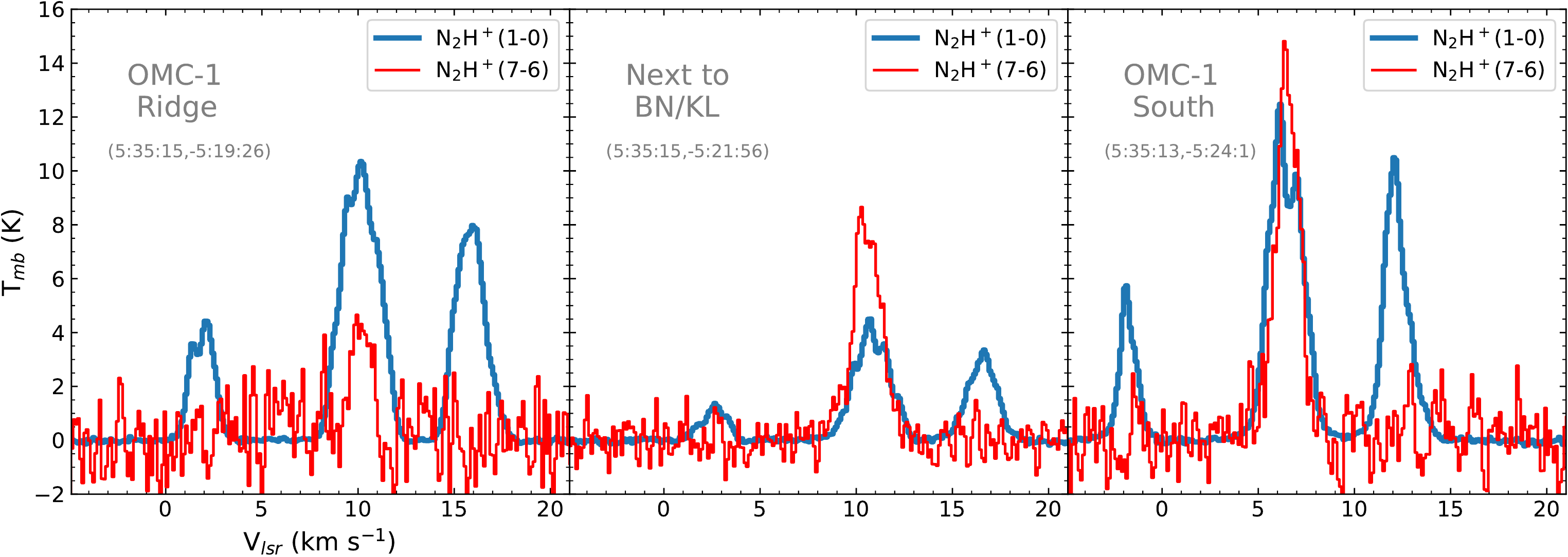}
	\caption{
		Representative N$_2$H$^+$~(1--0) (blue) and N$_2$H$^+$~(7--6) (red) spectra (in T$_{mb}$ units) found in the OMC-1 Ridge (left panel), the surroundings of the Orion BN/KL region (mid panel), and the OMC-1 South proto-cluster (right panel).
		The coordinates of our spectra are indicated in the upper left corner of each subplot (see also Fig.~\ref{fig:ratios}a).
	}
	\label{fig:spectra}
\end{figure*}

\section{Gas at densities $>10^7$~cm$^{-3}$ in OMC-1}\label{sec:highdensity}

In addition to their bright emission, the observed N$_2$H$^+$~(7--6) lines in OMC-1 show unprecedentedly high main-beam peak temperatures ($T_\mathrm{mb}$). As illustrated by several representative spectra shown in Fig.~\ref{fig:spectra}, we observe N$_2$H$^+$~(7--6) lines clearly detected well above the noise level our our data with $<rms> \sim 0.65$~K. We fitted all our N$_2$H$^+$~(7--6) spectra using a single Gaussian velocity component. 384 independent beams in our maps show N$_2$H$^+$~(7--6) spectra with signal-to-noise (S/N) larger than 3. Among them, the detected N$_2$H$^+$~(7--6)  emission presents mean values of $T_\mathrm{mb}$(N$_2$H$^+$~(7--6))~$\sim$~4~K and linewidths $\Delta V$(N$_2$H$^+$~(7--6))~$\sim$~1.3~km~s$^{-1}$ (Fig.\ref{fig:spectra} a \& b). Extremely bright spectra, with $T_\mathrm{mb}$(N$_2$H$^+$~(7--6))~$\sim$~10~K, are observed in the OMC-1 South region (Fig.\ref{fig:spectra}~c).
In many of these positions we find that the peak temperature of the N$_2$H$^+$~(7--6) transition (red spectra) matches and sometimes exceeds the corresponding peaks of their N$_2$H$^+$~(1--0) counterparts (blue spectra). 
At the OMC-1 ridge, comparisons between our SEPIA660 and ALMA data show values with $\frac{T_\mathrm{mb}(\mathrm{N}_2\mathrm{H}^+\mathrm{(7-6)})}{T_\mathrm{mb}(\mathrm{N}_2\mathrm{H}^+\mathrm{(1-0)})} \sim0.5$ while in the OMC-1 South and the surroundings of Orion BN/KL reach values above $\frac{T_\mathrm{mb}(\mathrm{N}_2\mathrm{H}^+\mathrm{(7-6)})}{T_\mathrm{mb}(\mathrm{N}_2\mathrm{H}^+\mathrm{(1-0)})} >1.0$ (see representative spectra).

We used the radiative transfer calculations provided by RADEX \citep{RADEX} to obtain the direct comparisons between the predicted N$_2$H$^+$ line intensities at different densities. Our calculations assume the latest collisional coefficients and energy levels provided by the Leiden Molecular Database \citep{LAMDA} without hyperfine structure.
We model the main properties of our lines adopting average line intensities and linewidth similar to those reported in our SEPIA660 observations (see above) for a characteristic column density of N(N$_2$H$^+$)~$=5\times10^{13}$~cm$^{-3}$ derived from detailed analysis of the hyperfine opacities obtained in our previous N$_2$H$^+$~(1--0) observations \citep[see Appendix B in][for a discussion]{HAC18}. 
Our models include three representative gas kinetic temperatures, namely, $T_\mathrm{K}=$~15~K, 25~K, and 35~K, describing the typical gas temperatures for the dense gas in OMC-1 consistent with previous temperature estimates \citep[see Fig.~\ref{fig:ratios}b;][]{HAC19}. Larger temperature values were not considered because the effective chemical destruction of N$_2$H$^+$ at $T_\mathrm{K}>$~35~K (see Sect.~\ref{sec:nebula}).

\begin{figure}
	\centering
	\includegraphics[width=\linewidth]{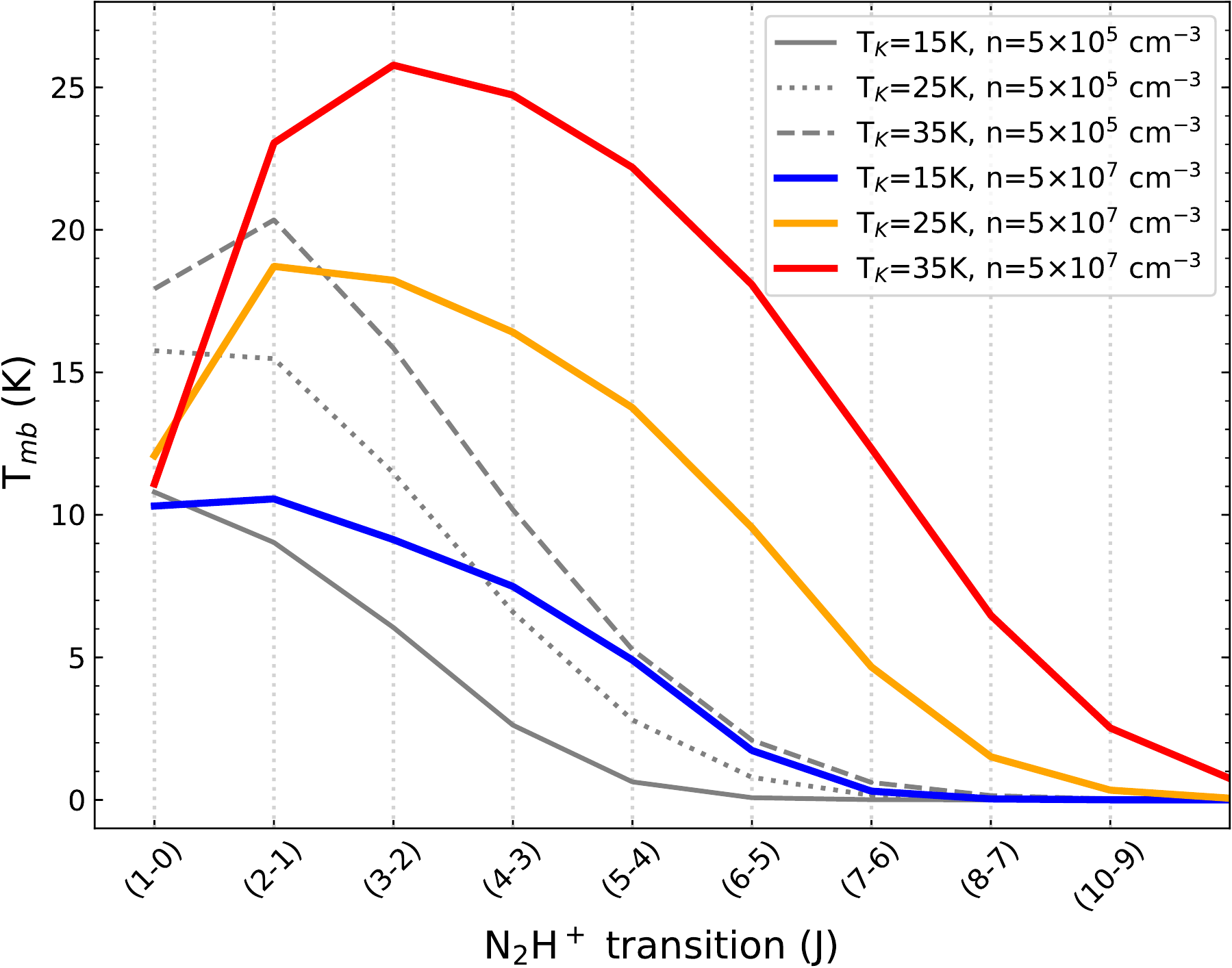}
	\includegraphics[width=\linewidth]{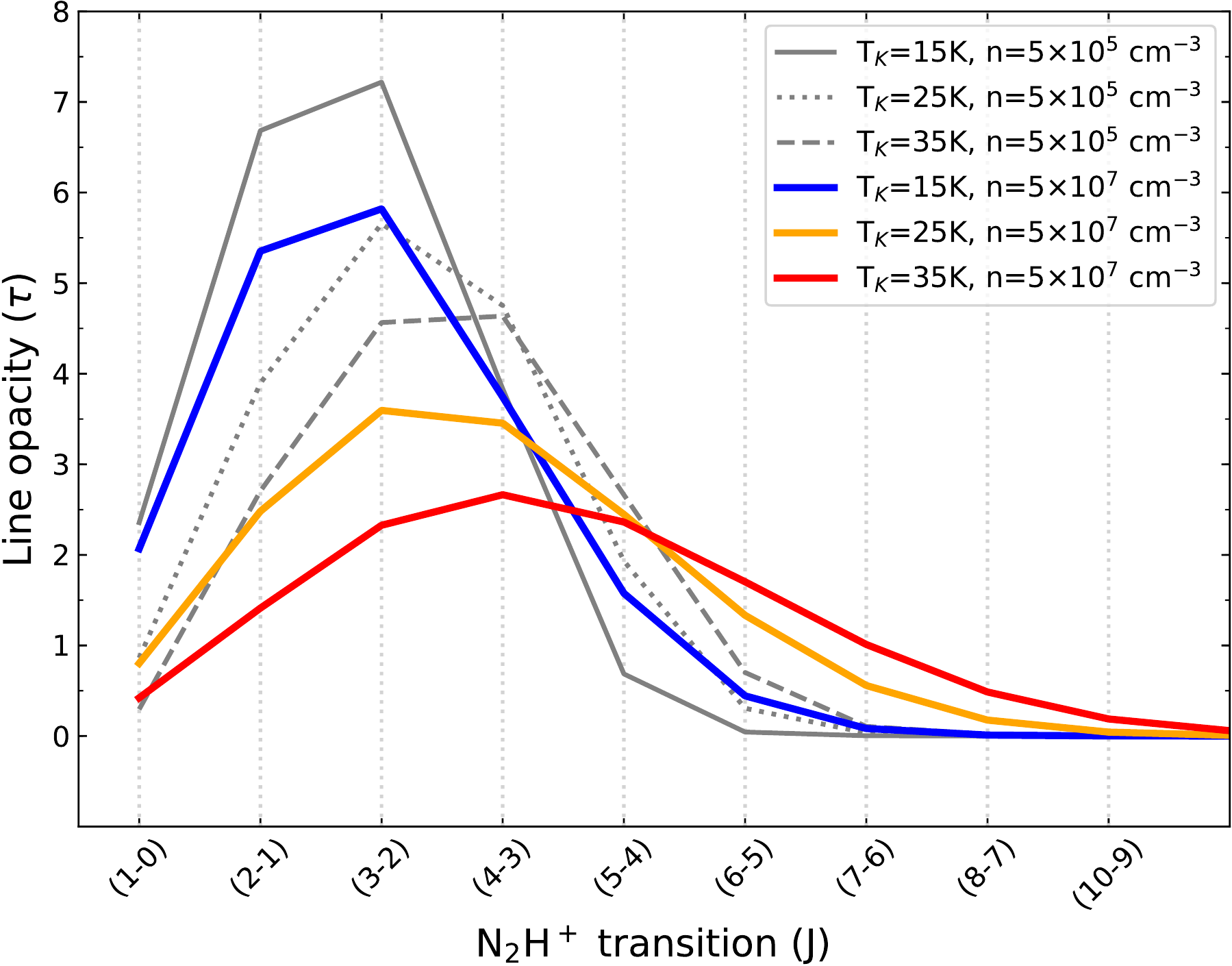}
	\caption{
	Expected  line peak temperatures $T_\mathrm{mb}$ {\bf (upper panel)} and line opacity $\tau $ {\bf (lower panel)} values for all N$_2$H$^+$ transitions (J~$\le$~10--9) predicted by our RADEX models  assumed as single-line components (aka without hyperfine structure).
	For simplicity, we represent only these models with densities  n(H$_2$)~$= 5\times 10^5$~cm$^{-3}$ (grey) and $5\times 10^7$~cm$^{-3}$ (colours), and temperatures of $T_\mathrm{K}=$~15~K, 25~K, and 35~K (see legend). 
	We notice how density significantly changes (aka skew) both $T_\mathrm{mb}$ and $\tau$ distributions towards high J-transitions. In this context, the detection of bright N$_2$H$^+$ (7--6) emission above $T_\mathrm{mb}>$~2~K can be used as direct probe of gas at densities above n(H$_2$)~$> 10^7$~cm$^{-3}$.
	}
	\label{fig:RADEX_Tpeak}
\end{figure}

\begin{figure}
	\centering
	\includegraphics[width=\linewidth]{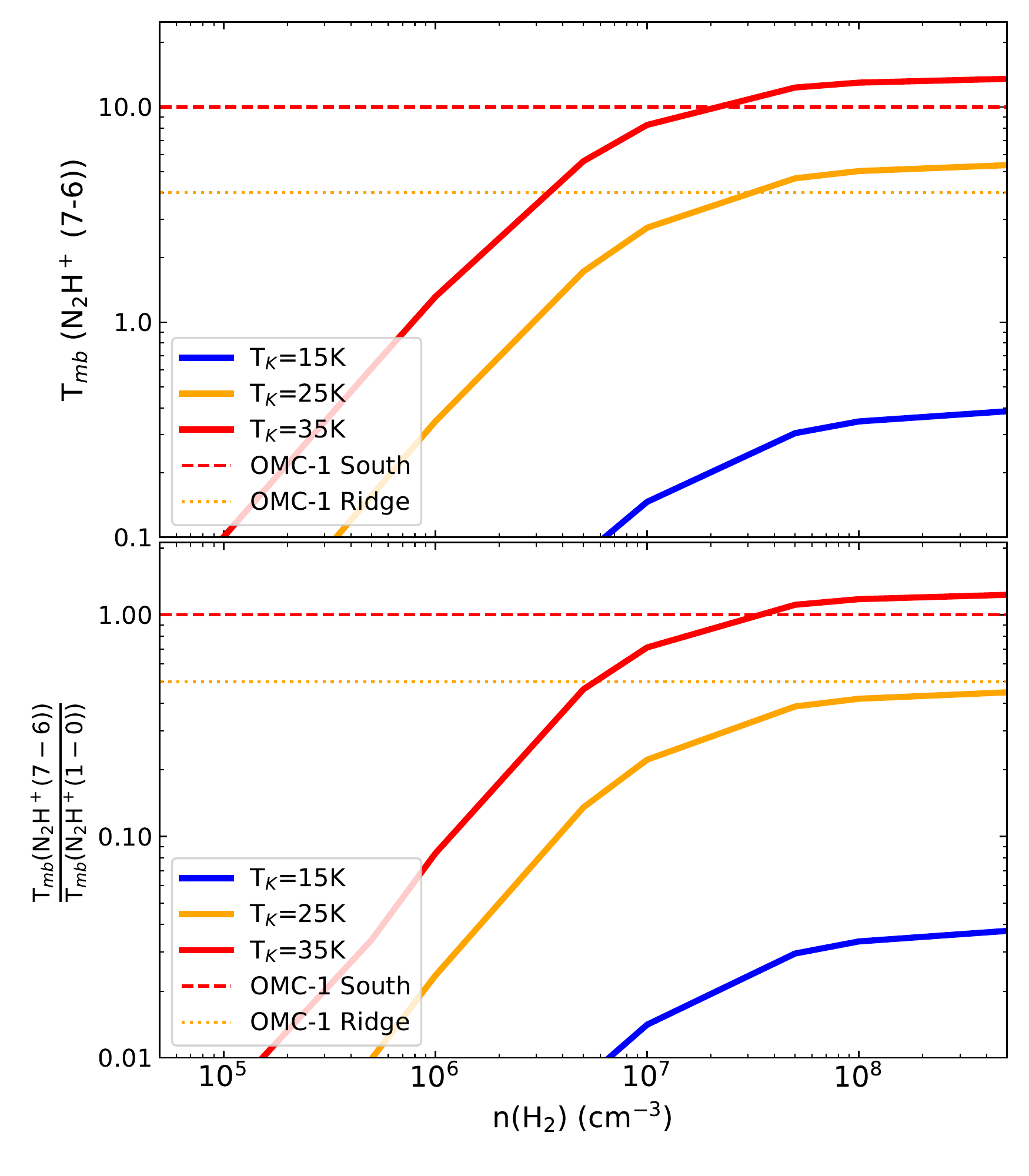}
	\caption{
		Expected N$_2$H$^+$~(7--6) line peak temperatures $T_\mathrm{mb}$(N$_2$H$^+$~(7--6)) ({\bf top panel}) and ratios respect the (1--0) transition  $\frac{T_\mathrm{mb}(\mathrm{N}_2\mathrm{H}^+\mathrm{(7-6)})}{T_\mathrm{mb}(\mathrm{N}_2\mathrm{H}^+\mathrm{(1-0)})}$
		({\bf bottom panel}) at densities n(H$_2$) between $10^5$~cm$^{-3}$ and $10^8$~cm$^{-3}$ and gas kinetic temperatures of T$_K=$~15~K (blue), 25~K (orange), and 35~K (red) predicted by our RADEX simulations \citep{RADEX}. The typical values found in OMC-1 South (red dashed line) and OMC-1 Ridge (orange dotted line) are indicated in both plots.
	}
	\label{fig:RADEX}
\end{figure}

Some caveats should be considered when interpreting our model results and their comparison with our observations. Most of these uncertainties are typically associated to the poorly characterized N$_2$H$^+$ (J~$\ge7-6$) transitions in comparison with its better known properties of lower J lines \citep[J~$\le 6-5$, e.g.][]{DAN05,PAG09}.
First, the absence of accurate estimates for the collisional coefficients for the N$_2$H$^+$ (7--6)  hyperfine transitions limits the depth of our analysis.
Thus, our models deliberately calculate the excitation conditions of N$_2$H$^+$  assuming a single line transition. With this approach we effectively add the contribution of all hyperfine components into a single line increasing the predicted N$_2$H$^+$ (1--0) and (7--6) line opacities and intensities up to a factor of $\sim$5-7  respect to their main hyperfine components. This choice is particularly justified in the case of our N$_2$H$^+$ (7--6) showing a compact hyperfine structure with almost all hypefine components blended within $\lesssim$~1.5~km~s$^{-1}$ (see Fig.~\ref{fig:spectra}).
Second, 
Our RADEX models adopt the LAMDA collisional rate coefficients that are extrapolated from those of HCO$^+$ \citep[see][]{FLO99}.
Recent collisional rate coefficients calculations include the hyperfine structure of  N$_2$H$^+$-H$_2$ up to J=7-6 \citep{LIQ15}. The difference between the two approximately introduces an uncertainty of a factor of 3-5 in our density estimates \citep{LIQ15}.
For consistency with our RADEX models, we fit our N$_2$H$^+$ spectra using a unique gaussian component. Comparisons with the peak line intensities predicted using the full hyperfine information (CDMS) indicate that our gaussian fits accurately reproduce the line peak temperatures of the observed N$_2$H$^+$ (1--0) and (7--6) spectra within $\le$~15\%. Linewidths, on the other hand, are affected by the superposition of multiple hyperfine components in both N$_2$H$^+$ (1--0) (central group) and (7--6) (full hyperfine structure) lines.
Our analysis focuses on a simplified description of the N$_2$H$^+$ (1--0) and (7--6) peak temperatures ($T_\mathrm{mb}$) meant to obtain a first order approximation to the gas densities along the ONC region. 
A more precise determination of the local gas densities in different positions of this cloud would require a more detailed treatment of the hyperfine structure and collisional coefficients of N$_2$H$^+$ \citep[e.g. see][]{KET10}  as well the simultaneous analysis of additional intermediate transitions such as (J=3--2) \citep{TEN20}.

The observed variations on the N$_2$H$^+$ line intensities can be understood from our RADEX models.
In Figure~\ref{fig:RADEX_Tpeak} we present the individual line peak temperatures T$_{mb}$  (upper panel) and opacities $\tau$ (lower panel) for all J-transitions (J$\le$10-9) considered in our RADEX models. For simplicity, we display only two characteristic density values, namely,  n(H$_2$)~$= 5\times10^5$~cm$^{-3}$ and n(H$_2$)~$= 5\times 10^7$~cm$^{-3}$, representing both low- and high-density regimes in our data, respectively.
Overall, higher temperatures and densities typically increase the excitation and emission of higher J-transitions. However, each of these variations show different behaviours along the N$_2$H$^+$ J-ladder.
As seen in Fig.~\ref{fig:RADEX_Tpeak} (upper panel), while temperature variations can potentially increase the peak temperatures in all J-transitions, only density is able to effectively excite high-J levels above J~$\ge$~6--5.
On the other hand, the opacity changes observed in Fig.~\ref{fig:RADEX_Tpeak} (lower panel) demonstrated how the combination of high temperatures and densities increases the population of high-J levels at expenses of those in lower J.
In high density and warm environments, the effective excitation of high-J levels J~$>$~6--5 is accompanied by a rapid reduction of the line opacities in all N$_2$H$^+$ transitions J~$\le$~4--3. 
While only  calculated for a single-component in our RADEX models, a similar variations of both line intensities and opacities with increasing densities are observed in radiative transfer calculations for the N$_2$H$^+$ (1-0) line including its entire hyperfine structure \citep[see Figure B.4 in][]{HAC18}.

Our previous plots demonstrate how the large energy difference between the N$_2$H$^+$ (1--0) ($E_u=$~4.7~K) and (7--6) ($E_u=$~125~K) lines provides crucial information about the densities of the star-forming gas in OMC-1.
In more detail, Figure~\ref{fig:RADEX} illustrates the line peak temperature $T_\mathrm{mb}$(N$_2$H$^+$~(7--6)) (top panel) and line ratios $\frac{T_\mathrm{mb}(\mathrm{N}_2\mathrm{H}^+\mathrm{(7-6)})}{T_\mathrm{mb}(\mathrm{N}_2\mathrm{H}^+\mathrm{(1-0)})}$ (bottom panel)  for densities between 10$^5$ and 10$^8$~cm$^{-3}$ predicted by our RADEX models.
For all temperatures our radiative transfer calculations show higher $T_\mathrm{mb}$(N$_2$H$^+$~(7--6)) values and $\frac{T_\mathrm{mb}(\mathrm{N}_2\mathrm{H}^+\mathrm{(7-6)})}{T_\mathrm{mb}(\mathrm{N}_2\mathrm{H}^+\mathrm{(1-0)})}$ ratios with increasing densities. The variations in both line peaks and ratios are primarily driven by the combination of both high densities and lukewarm temperatures required to effectively excite the N$_2$H$^+$~(7--6) transition (see Fig.~\ref{fig:RADEX_Tpeak}).
Even at relatively high temperatures, the detection of spectra showing $T_\mathrm{mb}$(N$_2$H$^+$~(7--6))~$>$~3~K guarantees the detection of gas of at densities n(H$_2$)~$> 10^6$~cm$^{-3}$. For the same detection threshold our models predict larger densities for decreasing temperatures (see coloured lines in the plots) or column densities (not shown). 

The use of RADEX radiative transfer models allows us to constrain the gas densities traced by our new SEPIA660 observations.
The $T_\mathrm{mb}$(N$_2$H$^+$~(7--6)) and $\frac{T_\mathrm{mb}(\mathrm{N}_2\mathrm{H}^+\mathrm{(7-6)})}{T_\mathrm{mb}(\mathrm{N}_2\mathrm{H}^+\mathrm{(1-0)})}$ values detected along the OMC-1 region (see also horizontal lines in Fig.~\ref{fig:RADEX}) rule out densities below n(H$_2$)~$<10^6$~cm$^{-3}$ and temperatures $T_\mathrm{K}<$~20~K. Instead, the observed high values for both peak temperatures and line ratios detected can only be reproduced if the gas detected in N$_2$H$^+$ (7--6) is at densities n(H$_2$)~$> 5\times 10^7$~cm$^{-3}$. Even higher densities, with n(H$_2$)~$\sim 10^8$~cm$^{-3}$, would be also consistent with the detected line ratios in the OMC-1 South proto-cluster. Secondary differences between these two regions can be attributed to the slightly warmer conditions found towards OMC-1 South ($T_\mathrm{K}\sim$~35~K) compared to the OMC-1 Ridge ($T_\mathrm{K}\sim$~25~K) \citep{HAC19}.

The unique detection of N$_2$H$^+$~(7--6) transition  
unambiguously demonstrates the presence of gas at ultra-high densities in OMC-1. Previous calculations based on the N$_2$H$^+$~(1--0) line opacities \citep{HAC18} and the N$_2$H$^+$~(3--2) intensities \citep{TEN20} derived density values between n(H$_2$)~$=10^6 - 10^7$~cm$^{-3}$. The inclusion of this new N$_2$H$^+$~(7--6) transition potentially increases the density estimates by at least a factor of 5 in regions such as the OMC-1 South. On the other hand, our new SEPIA observations demonstrate that part of the dense material traced in N$_2$H$^+$ is effectively heated by the ONC Nebula at temperatures above $T_\mathrm{K}\gtrsim$~30~K. Previous estimates derived temperatures for the dense gas traced in  N$_2$H$^+$ of $T_\mathrm{K}=$~20~K \citep[see][]{TEN20}. Our radiative transfer calculations indicate that significant fractions of this dense material are consistent with temperatures of $T_\mathrm{K}\gtrsim$~30~K.

The presence of large amounts of N$_2$H$^+$ at high temperatures appears to be counter-intuitive. CO is expected to be evaporated from the dust grains at $T_\mathrm{dust}>$~15~K \citep{BER07}.  Previous observations report dust effective temperatures ($T_\mathrm{dust}$) similar to the gas kinetic temperatures along OMC-1 showing differences of  $|T_\mathrm{dust}- T_\mathrm{K}|\lesssim $~5~K \citep[see][]{HAC19}\footnote{The effective dust temperature ($T_\mathrm{dust}$) is usually obtained from a single-component black-body fit of the observed FIR luminosities \citep[e.g.][]{LOM14}. This effective temperature describes the average dust grain temperature ($T_\mathrm{grain}$) weighted along the line-of-sight. Biased towards warmer temperatures producing a bright FIR emission, the effective dust temperature typically overestimates the local dust grain temperatures in cold and dense (aka well-shielded) regions showing fainter FIR emission, that is, $T_\mathrm{dust}\ge T_\mathrm{grain}$.}.
At the $T_\mathrm{dust}\sim T_\mathrm{K}=$~20-30~K observed along the ONC region (see Fig.~\ref{fig:ratios}b), CO is then expected to quickly destroy N$_2$H$^+$ via reaction (\ref{d1}) once CO is back into the gas phase. 
This evaporation process could be counterbalanced by the short freeze-out timescales of this molecule ($\sim$~100~yr) expected at the ultra high gas densities detected in this region \citep[$\tau_{f-o}\sim 5\times 10^9 / \mathrm{n}(\mathrm{H}_2)$ yr, see][]{BER07} continuing operating at high temperatures \citep[see Appendix B in][]{HAR15}. 
If mixed with other ices, CO could also disorb at higher temperatures delaying its evaporation from the dust grains \citep{VIT04}.
Moreover, the presence of N$_2$H$^+$ could be enhanced by the reduction of the dissociative recombination rate of reaction (\ref{d2}) at high temperatures \citep{VIG12}.
The combination of these effects appear to favour the survival of N$_2$H$^+$ at lukewarm temperatures 20~K$\lesssim T_\mathrm{K}\lesssim$~35~K in extremely dense environments such as the surroundings of the ONC.

Based on our ALMA measurements, we estimate a minimum of 30~M$_\odot$ at densities n(H$_2$)~$>10^7$~cm$^{-3}$ within our maps. Our calculations include those positions with significant emission in N$_2$H$^+$~(7--6) at S/N~$\ge$~3. Due to the excitation conditions of this line, our selection criteria restrict these mass estimates to dense gas pockets at temperatures above $T_\mathrm{K}>$~25~K (see Fig.\ref{fig:spectra}). According to our (1--0) detections, larger mass reservoirs are likely present at lower temperatures towards the north and west of the OMC-1 region (e.g.  see N$_2$H$^+$~(1--0) maps in Fig.~\ref{fig:observations}). This conclusion is reinforced by the bright and extended N$_2$H$^+$~(3--2) emission detected towards the entire OMC-1 region \citep{TEN20}. Our mass estimates should therefore be considered as lower limits of total amount of gas at ultra-high densities in this region. 

\section{Star-formation at extremely high densities}\label{sec:final}

The widespread detection of N$_2$H$^+$~(7--6) emission shown in our observations (Sect.~\ref{sec:observations}) illustrates the extreme physical conditions of the gas in young massive clusters such as the ONC.
Our  Early Science SEPIA660 observations demonstrate the existence large volumes of gas at densities n(H$_2$)~$>10^7$~cm$^{-3}$ in close proximity to the ONC (Sect.~\ref{sec:nebula}).  These densities are at least two orders of magnitude larger than those gas densities found in the densest cores in low-mass star-forming regions, typically with n(H$_2$)$\sim10^5$~cm$^{-3}$ \citep[e.g.][]{CAS02} (Sect.~\ref{sec:highdensity}). 
Previously suggested to be restricted to the Orion BN/KL hot core \citep{GOD11}, our new observations extend the presence of lukewarm gas at densities above n(H$_2$)~$>10^7$~cm$^{-3}$ at scales of approximately 1~pc. 

These results explain the extraordinary star-formation properties found along the OMC-1 Ridge and the OMC-1 South proto-cluster. Recent millimeter continuum and X-ray surveys found a typical separation between young embedded sources of 2000 AU \citep[OMC-1 Ridge,][]{TEI16} and 600 AU \citep[OMC-1 South,][]{RIV13}.  
These values are in excellent agreement with the corresponding Jeans fragmentation lengths ($\lambda _J=\frac{c_s}{G \mathrm{n(H}_2\mathrm{)} }$) for a gas at temperatures of $T_\mathrm{K}=$~30~K (i.e. $c_s=0.35$~km~s$^{-1}$) at densities between n(H$_2$)~$=10^7$~cm$^{-3}$ ($\lambda_J \sim 1450$~AU) and n(H$_2$)~$=10^8$~cm$^{-3}$ ($\lambda_J \sim 450$~AU).
With expected free-fall times of $\tau_{ff}\lesssim10^4$~yrs, these densities are also consistent with the young ages and high SFRs found in the embedded populations in regions such as OMC-1 South \citep[see][]{RIV13}. 
Moreover, our SEPIA660 results also confirm the density values predicted for the star-forming fibers found in the OMC-1 region by \citet{HAC18}. Compared to those low-mass Herschel filaments showing widths of $\sim$~0.1~pc \citep{ARZ11}, the reported densities of n(H$_2$)$>10^7$~cm$^{-3}$ explain the much narrower fibers widths of $\sim$~0.03~pc found in this massive ONC region \citep[see][for a discussion]{HAC18}. 

The unusually large volume densities of the gas found in OMC-1 illustrate the extraordinary properties of this cloud.
The high densities detected in fibers and cores (traced in N$_2$H$^+$) allow these structures to survive the strong radiative and mechanical feedback produced by the O-type stars in the Trapezium shielded behind large column densities of warm molecular material (observed in C$^{18}$O). 
Still, two competing mechanism operate in this cloud.
First, the reported $T_\mathrm{K}>$~20~K values in the surroundings of the ONC 
indicate that some of these structures could be photoevaporated by the HII nebula in relative short timescales $\tau_{photo}$. On the other hand, this destruction process is counteracted by the rapid $\tau_{ff}$ collapse of the ultra dense star-forming gas revealed by our N$_2$H$^+$ (7--6) observations (see above).
The detection of large number of young embedded sources in regions like OMC-1 South \citep[e.g.][]{RIV13} indicates that $\tau_{photo} \gg \tau_{ff}$ even under these extreme gas conditions.
In agreement to recent simulations \citep{DAL14}, our observations suggest that feedback may have little effect on the evolution of the gas at extremely high densities found in this massive cluster. 

The unique combination of sensitivity and mapping capabilities of the new APEX-SEPIA660 receiver opens a new window for ISM studies at high frequencies.
These ultra high gas densities reported in OMC-1 mimic the physical conditions of more distant and massive environments such as the Central Molecular Zone or Starburst Galaxies. Our new SEPIA660 observations reveal this OMC-1 cloud as unique laboratory to investigate the fragmentation, collapse, and chemical evolution of the gas at extreme density conditions with unprecedented detail. 
Moreover, the confirmed detection of bright and extended emission of N$_2$H$^+$~(7--6) (652~GHz) offers the possibility of observing these regions at ultra high resolutions with ALMA (Band 9).

\begin{acknowledgements}
	The APEX SEPIA receiver is a joint development by the Group of Advanced Receiver Development (GARD, Gothenburg, Sweden) from the Onsala Space Observatory (OSO, Sweden), the Netherlands Research School for Astronomy (NOVA, The Netherlands), and the European Southern Observatory (ESO).
	D. H. acknowledges support from the EACOA fellowship from the East Asian Core Observatories Association.
	This paper makes use of the following ALMA data: ADS/JAO.ALMA\#2015.1.00669.S. ALMA is a partnership of ESO (repre- senting its member states), NSF (USA) and NINS (Japan), together with NRC (Canada) and NSC and ASIAA (Taiwan) and KASI (Republic of Korea), in cooperation with the Republic of Chile. The Joint ALMA Observatory is operated by ESO, AUI/NRAO and NAOJ. Based on observations carried out with the IRAM 30m Telescope. IRAM is supported by INSU/CNRS (France), MPG (Germany) and IGN (Spain).
	This research made use of APLpy, an open-source plotting package for Python \citep{Astropy}. This paper made use of the TOPCAT software \citep{TAY05}.

\end{acknowledgements}

%
%

\end{document}